# Tritium Separation from Gaseous $^{1,2,3}$H Isotopologue Mixtures by Selective Adsorption on Ag-Exchanged Zeolite Type Y

Alexandra Becker*[a, b], Holger Lippold[a], Michael Hirscher[c, d], and Cornelius Fischer[a, b]

[a]    A. Becker, Dr. H. Lippold, Prof. Dr. C. Fischer
       Institute for Resource Technology, Department of Reactive Transport
       Helmholtz-Zentrum Dresden-Rossendorf
       Permoserstraße 15, 04318 Leipzig, Germany
       E-mail: alexandra.becker@kit.edu

[b]    A. Becker, Prof. Dr. C. Fischer
       Faculty of Chemistry
       Leipzig University
       Johannisallee 29, 04103 Leipzig, Germany

[c]    Dr. M. Hirscher
       Max Planck Institute for Intelligent Systems
       Heisenbergstraße 3, 70569 Stuttgart, Germany

[d]    Dr. M. Hirscher
       Advanced Institute for Materials Research (WPI-AIMR)
       Tohoku University
       Sendai, 980-8577, Japan

**Abstract:** Efficient separation of hydrogen isotopologues is crucial for applications such as the recycling of exhaust streams in nuclear fusion reactors. Using thermal desorption spectroscopy (TDS), we report, for the first time, on separation experiments of a ternary $^{1,2,3}$H isotope mixture, achieving tritium enrichment of 70% from an initially equimolar (1:1:1) gas mixture, based on selective adsorption using an Ag(I)-exchanged zeolite type Y. Further experiments on binary hydrogen isotope mixtures validated numerical predictions of the separation efficiency for $T_2$. Specifically, the high selectivity for tritium over protium of 38.2 makes the Ag(I)-exchanged zeolite an excellent candidate for energy-efficient isotope separation at liquid-nitrogen temperature.

Among the $CO_2$-free energy options, nuclear fusion is regarded as particularly promising and is being pursued in large-scale projects such as ITER.[1] The reaction fuel consists of the hydrogen isotopes deuterium and tritium. However, only about 2% of the fuel is consumed and therefore, a continuous recycling by isotope separation is required.

Currently, tritium is used in self-powered illumination technologies, for instance in watches and aviation instruments. In these applications, ultra-thin glass capillaries are filled with tritium gas, where β-electrons from nuclear decay excite a luminescent coating to produce light.[2] Recycling and reuse of such devices necessitate efficient purification and separation processes, particularly to recover tritium from $H_2$.

Cryogenic distillation is an established method for separating the stable isotopes hydrogen and deuterium.[3] However, analyses indicate that this technique is not well suited for the broad range of compositions expected in fusion reactor processes.[4] In recent years, alternative ways have been developed, particularly adsorption-based separation using microporous materials such as metal-organic frameworks (MOFs) or zeolites.

In microporous materials, two quantum effects can be exploited for the separation of hydrogen isotopologues: kinetic quantum sieving (KQS) and chemical affinity quantum sieving (CAQS), both governed by the pore structure and chemical composition of the adsorbent material.[5] KQS, first described by Beenakker et al. in 1995,[6] arises from differences in the effective particle size (de Broglie wavelength) of the isotopologues, resulting in preferential adsorption of the heavier isotope.[7,8] A variety of crystalline materials with chemically tunable cavities are currently being investigated to elucidate the KQS mechanisms and to optimize separation efficiency.[9] CAQS, in contrast, relies on strong adsorption sites.[10,11] The isotopologues exhibit distinct adsorption enthalpies, determined by the mass-dependent zero-point energy (ZPE) in the van der Waals potential, which allows for quantum sieving on strong binding sites even at temperatures above 80 K.[12-14]

Various classes of porous materials have been experimentally investigated for the separation of the stable hydrogen isotopes $H_2$ and $D_2$.[12,15-17] However, temperature-resolved studies on $T_2$ are limited and separation factors have so far, mostly, been predicted theoretically, based on calculations consistent with experimental $H_2/D_2$ separation data. The calculations predict selectivities of $S_{T/H}$ = 40.6 and $S_{T/D}$ = 3.5 for 1:1 binary gas mixtures at 80 K on Ag-exchanged ZSM-5 zeolite,[18] with selectivities based on the molar ratios of the adsorbed amounts.

For radioactive tritium, the structural stability of nanoporous materials needs a critical consideration. Zeolites are particularly promising for tritium separation, having demonstrated structural resilience under varying radiation doses and tolerance to tritium decay.[19] Open questions remain regarding the radiolytic resistance of other nanoporous materials considered for separation applications, such as MOFs,[20] as well as of two-dimensional layered materials like graphene, particularly with respect to structural modifications.[21]

Thermal desorption spectroscopy (TDS)[5] is the method of choice for investigating the adsorption and desorption of hydrogen isotopologues under temperature-controlled conditions. It enables direct sample activation within the setup and operates over a wide temperature range, starting at cryogenic conditions. To date, TDS studies involving tritium have not been conducted, as handling highly radioactive gases requires specialized analytical solutions that comply with safety regulations (see Supporting Information).

For the first time, we validate the predicted separation factors using ternary hydrogen isotope mixtures that include tritium by employing the TDS technique with a setup similar to Zhang et al.[14] This allows for a direct comparison with previous studies on protium-deuterium mixtures.

As a microporous adsorbent substrate, an Ag(I)-exchanged zeolite type Y (AgY)[14] was used (see Supporting Information). A 1.7 mg sample was activated at 500 K for 2 h prior to the adsorption experiment. The sample was cooled to 82 K under vacuum using liquid nitrogen. After exposure to pure gases or isotope mixtures, desorption of the isotopes on heating up to 300 K was monitored by TDS using a quadrupole mass spectrometer. Calibration for $H_2$ and $D_2$ was performed via weight determination using a $Pd_{95}Ce_5$ alloy. For $T_2$, desorbed amounts were determined via liquid scintillation counting after catalytical conversion to HTO (CuO catalyst, 910°C) and were correlated to the TDS data integral, eliminating exposure to the highly radioactive gas. Single-isotope analyses showed highly reproducible uptakes of 0.35 mmol/g ($D_2$) and 0.33 mmol/g ($T_2$), whereas protium uptake was more than twice as high at 0.72 mmol/g. The data for the single gas measurements can be found in the Supplementary Information (Fig. S2, Fig. S3 and Table S1). The selectivities for the binary mixtures were calculated from the individual gas uptakes as determined from the desorption spectra.

Of particular interest is the separation of isotopologues in a ternary mixture relevant for fusion power plants,[22] which we have now experimentally realized, as shown in Fig. 1. The TDS spectra reveal pronounced selectivities, evident from the total amounts desorbed after exposure to an equimolar $H_2/D_2/T_2$ mixture and removal of the supernatant gas at 83 K. Tritium adsorption is notably favored, whereas deuterium adsorption is moderate and protium adsorption is almost completely suppressed. The resulting percentages are 2.54% for protium, 27.81% for deuterium and 69.65% for tritium, when solely taking the homonuclear isotopologues into account. All gas compositions including the estimated values for the heteronuclear isotopologues are given in the Supplementary Information (Table S2).

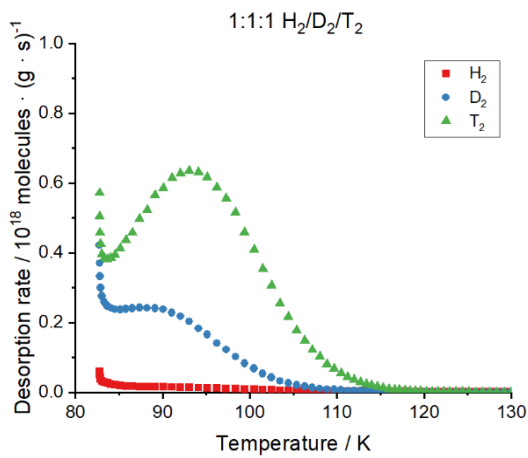

**Figure 1.** Desorption spectra of a ternary 1:1:1 hydrogen isotopologue mixture after adsorption at ~ 83 K, introduced at a total pressure of 1 kPa.

The $D_2$ measurements by MS were minimally affected by HT formation, an effect that proved negligible (see Table S2, Supporting Information). The total gas uptake was 0.27 mmol/g, well below surface saturation.[23]

TDS analysis of the constituent binary isotopologue mixtures provides insight into the thermodynamic basis of the superimposed selectivities observed in the ternary system. Both the magnitude of gas uptake and the temperature corresponding to maximum desorption rates are to be considered to characterize the system's selective behavior. For an equimolar $H_2/D_2$ mixture (Fig. 2A), the total uptake was 0.32 mmol/g. The $D_2$ signal is at maximum at ~ 92 K, while the $H_2$ curve shows no maximum within the measured temperature range. The corresponding selectivity $S_{D/H}$ was calculated as 7.7, in agreement with previously reported experimental data.[14] This enables direct quantitative comparison with tritium-containing binary mixtures: For the $D_2/T_2$ mixture (Fig. 2B), a plateau is observed at ~ 89 K for $D_2$ and a maximum at ~ 95 K for $T_2$. The higher desorption temperature of $T_2$ reflects its elevated adsorption enthalpy and correlates with its preferential adsorption. Shifts in the maximum desorption temperatures, as observed for $D_2$ in the presence of $H_2$ and $T_2$, indicate the influence of competitive adsorption effects. The obtained selectivity of $S_{T/D}$ = 2.5, with a total gas uptake of 0.22 mmol/g at a slightly higher temperature of ~ 85 K, is comparable to the predicted value ($S_{T/D}$ = 3.5) for the separation on Ag-exchanged ZSM-5.[18] This selectivity is very close to a previous analysis using Ag(I)-exchanged zeolite type Y ($S_{T/D}$ = 2.44 at 77 K, uptake ratio of tritium to deuterium 1.73),[23] where an indirect approach was employed for analyzing the gas mixture composition in a reservoir, rather than directly monitoring the adsorption and desorption behavior of the substrate as a function of the temperature. Questions regarding isotope exchange remained unresolved, and the method offers limited or no access to desorption kinetics, competitive effects and ternary gas mixtures.

For the binary $H_2/T_2$ mixture (Fig. 2C), the total gas uptake was 0.28 mmol/g, with $H_2$ contributing minimally. Consistent with previous observations, no desorption maximum for $H_2$ was detected within the measured temperature range, whereas $T_2$ exhibits a maximum at ~ 91 K. The corresponding selectivity $S_{T/H}$ was calculated as 38.2, closely matching predictions. Xiong et al. reported a selectivity of 40.6 for $T_2$ over $H_2$ on Ag-ZSM-5 at 80 K.[18]

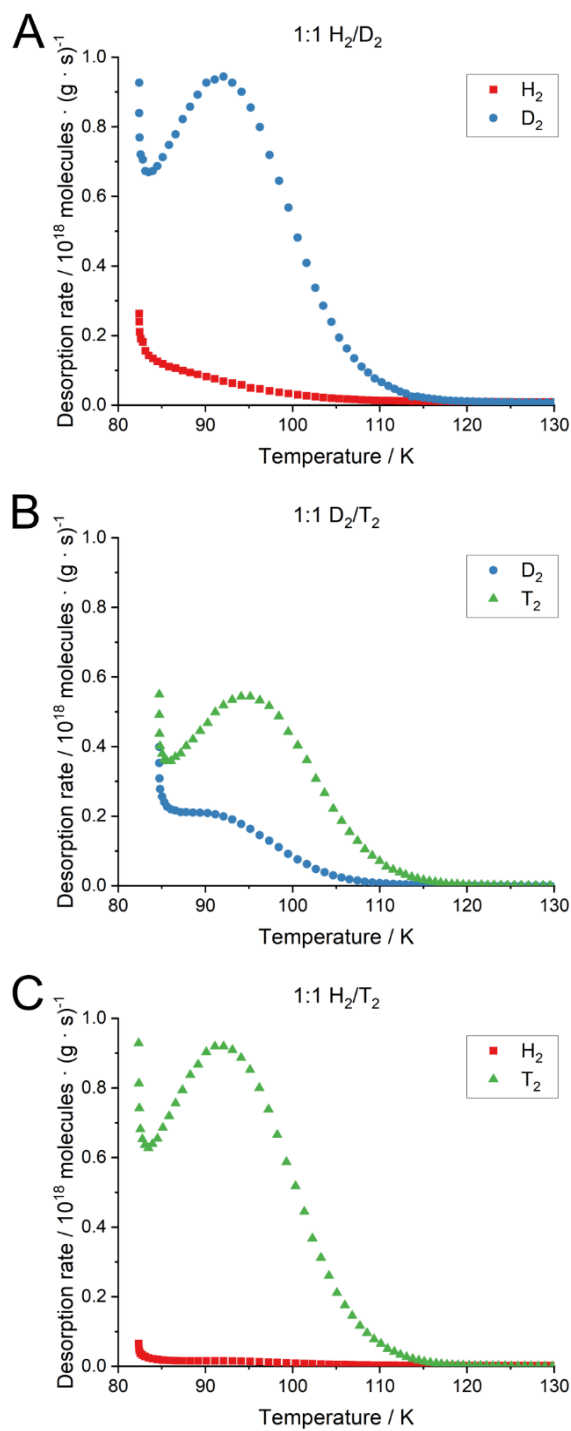

**Figure 2.** Desorption spectra of the binary mixtures introduced at a total pressure of 1 kPa. A: protium and deuterium (adsorption at ~ 82 K), B: deuterium and tritium (~ 85 K), C: protium and tritium (~ 82 K).

For all gas mixtures, the extent of isotope exchange was found to be minimal, despite both adsorption and radiolytic reactions. TDS measurements showed only very weak signals of HD, HT and DT, indicating that isotope exchange occurs at a very low rate (Supporting Information, Table S2).

We present an efficient method for separating tritium from binary and ternary isotope mixtures by selective adsorption on an Ag(I)-exchanged zeolite Y. This straightforward process exploits the stronger interaction of hydrogen isotopologues with $Ag^+$ cations in the micropores. Owing to the strong interaction, the required cryogenic temperatures can be achieved using liquid nitrogen, which offers a significant energy advantage over conventional methods, such as cryogenic distillation. For binary, equimolar isotope mixtures, high selectivities of 7.7, 38.2, and 2.5 have been determined for $D_2/H_2$, $T_2/H_2$, and $T_2/D_2$, respectively, achieved by chemical affinity quantum sieving on $Ag^+$ sites. Furthermore, a minimal isotope exchange is observed and the zeolite is radiation resistant making it a potential material for efficient tritium enrichment in large-scale technological facilities.

**Supporting Information**

The authors have cited additional references within the Supporting Information.[24,25]

**Acknowledgements**

We are very thankful to Wolfgang Schmidt and Florian Baum (Max-Planck-Institut für Kohlenforschung, Mülheim an der Ruhr, Germany) for preparing the Ag(I)-exchanged zeolite Y samples. Special thanks go to Bernd Ludescher for constructing and building the new TDS device. We also thank Linda Zhang and Prantik Sarkar for their help and knowledge with the experimental work and data evaluation.

This project is funded by the Deutsche Forschungsgemeinschaft (DFG), Project-ID 443871192, GRK 2721 "Hydrogen Isotopes $^{1,2,3}$H".

**Keywords:** microporous materials • zeolites • deuterium • tritium • hydrogen isotopologue separation

**Entry for the Table of Contents**

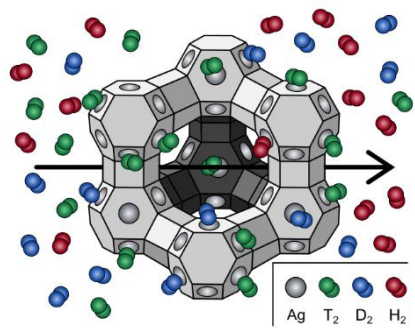

We report a temperature-resolved tritium enrichment in Ag-exchanged zeolite type Y using thermal desorption spectroscopy (TDS). The material enables selective enrichment of tritium from hydrogen isotope mixtures. These results highlight both the zeolite and the separation approach as promising candidates for applications involving gaseous tritium.

Supporting Information

# Tritium Separation from $^{1,2,3}$H Isotopologue Mixtures by Selective Adsorption on Ag-Exchanged Zeolite Type Y

Alexandra Becker*[1, 2], Holger Lippold[1], Michael Hirscher[3,4], Cornelius Fischer[1, 2]

[1]Helmholtz-Zentrum Dresden-Rossendorf, Institute of Resource Ecology, Department of Reactive Transport, Leipzig, Germany

[2]Leipzig University, Faculty of Chemistry, Leipzig, Germany

[3]Max Planck Institute for Intelligent Systems, Stuttgart, Germany

[4]Advanced Institute for Materials Research (WPI-AIMR), Tohoku University, Sendai, Japan

## Table of Contents

## 1. Materials and Methods

For the separation experiments, Ag(I)-exchanged zeolite type Y (AgY) was used, which was prepared at Max-Planck-Institut für Kohlenforschung (Mülheim an der Ruhr, Germany) using an ion exchange reaction on NaY with $AgNO_3$.[1] It was characterized with powder XRD. The studied sample was activated under vacuum in the TDS at 500 K for 2 h to desorb $H_2O$ and other components. The TDS used is a similar setup to the device used by Zhang et al.[1] Cooling of the sample was realized with liquid nitrogen.

Access to $T_2$ was provided through a stainless steel vacuum compartment system (RC TRITEC AG, Switzerland) using the uranium getter technology,[2] which is based on reversible thermal detritiation of a $UT_3$ source (3.7 TBq). $T_2$ gas is set free at a temperature of ~ 500°C and is completely rebound by chemisorption at room temperature. Before $T_2$ dosing, the decay product $^3He$ was baked out of the uranium bed in a heating-cooling sequence.

To calibrate the quadrupole mass spectrometer (PrismaPro QMG 250, Pfeiffer Vacuum, Germany) used for monitoring gas desorption, a $Pd_{95}Ce_5$ alloy was loaded with known masses of pure $H_2$ or pure $D_2$.[1] As for tritium, a different method had to be used, because weighable amounts cannot be safely handled due to the very high radioactivity. Instead, tritium desorbed in a TDS run was purged in an air flow from the outlet of the MS pump to a reactor tube where it was converted to HTO in a combustion reaction with $O_2$ over CuO catalyst (Merck, Germany) at 910°C. The reaction product was collected in a water trap, which was sampled for analysis by liquid scintillation counting (LSC) using a Tri-Carb 3110 TR Liquid Scintillation Analyzer (Perkin Elmer, US) and Ultima Gold™ scintillation cocktail (Perkin Elmer, US). The completeness of conversion and trapping was recently demonstrated.[3] The determined $T_2$ amount was correlated with the TDS peak integral to calibrate the MS signal.

In preparation of the separation experiments, 1.7 mg zeolite was exposed to all pure isotopologue gases to see how the individual gases are ad- and desorbed. For this, the sample chamber was flooded with 1 kPa of the gases at room temperature. Then the sample was cooled down to temperatures as low as approx. 82 K. After waiting an additional 10 min, the supernatant gas was removed by evacuation (in the case of $H_2$ and

D$_2$) or by sorption onto a second uranium bed (in the case of T$_2$). To quantify the adsorbed gas, a temperature program was run while simultaneously measuring the desorbed gas with the MS. The program started at the exposure temperature and was run up to 300 K at a heating rate of 0.1 K/s. These measurements were done twice for each isotopologue.

The process for the separation experiments was similar, but the sample was cooled first and then exposed to the binary or ternary gas mixture for 10 min. Again, a total pressure of 1 kPa was applied (as sum of the partial pressures of the gas components), and the supernatant gas was removed by evacuation or sorption onto uranium depending on whether tritium was part of the mixture. All gas mixtures were prepared immediately before the separation experiment. To calculate the selectivities of the binary gas mixtures, the uptake ratio of the heavier isotope to the lighter isotope was determined from each desorption spectrum.

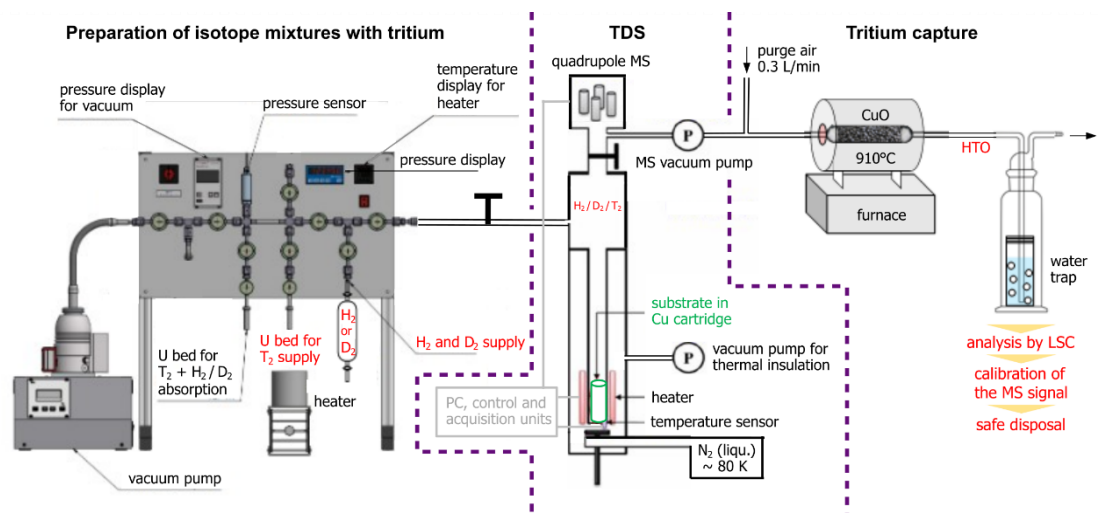

**Figure S1:** Schematic of the experimental setup. Left: the stainless steel vacuum compartment system for tritium distribution and preparation of hydrogen isotope mixtures, middle: the TDS setup with the quadrupole mass spectrometer, right: the tritium capture setup with flow reactor.

## 2. Additional data

The results of two sets of pure gas measurements are shown in Fig. S2 and Fig. S3. The uptake for deuterium and tritium is 0.35 mmol/g and 0.33 mmol/g, respectively. For hydrogen, the total uptake was significantly higher at 0.72 mmol/g. The desorption maxima are very similar, slightly shifted to higher temperatures for the heavier isotopes, at ~ 88 K ($H_2$), ~ 91 K ($D_2$) and ~ 91 K ($T_2$). This is in accordance with the higher heat of adsorption for the heavier isotopes.

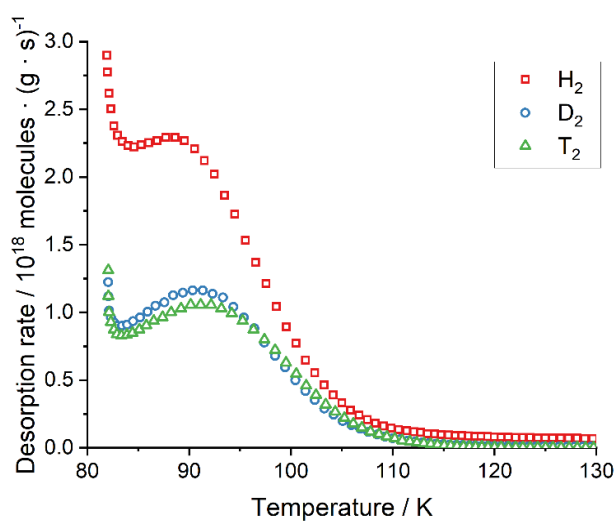

**Figure S2:** Desorption spectra for the individual gases after adsorption at ~ 82 K, introduced at a pressure of 1 kPa.

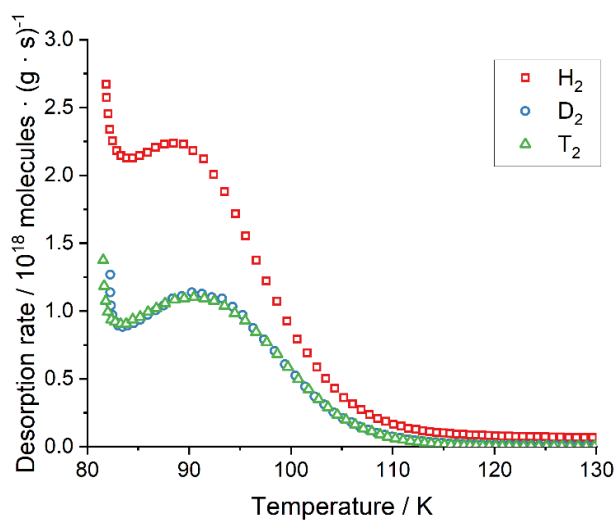

**Figure S3:** Desorption spectra for the individual gases after adsorption at ~ 82 K, introduced at a pressure of 1 kPa (repetition experiments).

**Table S1:** Overview of gas uptakes and the approximate maximum positions for all TDS experiments with the individual gases.

|  | First measurement | | Second measurement | |
| --- | --- | --- | --- | --- |
| Gas | Total uptake (mmol/g) | Maximum (K) | Total uptake (mmol/g) | Maximum (K) |
| $H_2$ | 0.72 | 88.2 | 0.71 | 88.5 |
| $D_2$ | 0.35 | 90.8 | 0.34 | 90.3 |
| $T_2$ | 0.33 | 91.1 | 0.34 | 90.5 |

**Table S2:** Percentages of desorbed gas components taken from the desorption spectra including the mixed isotopologues. The number of desorbed molecules for HD, HT and DT was estimated using the average of the calibration constants of the pure gases. The $D_2$ amount measured for the ternary mixture includes a small unknown amount of HT.

| Mixture | $H_2$ | HD | HT | $D_2$ | DT | $T_2$ |
| --- | --- | --- | --- | --- | --- | --- |
| $H_2/D_2$ | 11.16% | 2.36% | - | 86.47% | - | - |
| $H_2/T_2$ | 2.51% | - | 1.51% | - | - | 95.98% |
| $D_2/T_2$ | - | - | - | 27.69% | 3.72% | 68.58% |
| $H_2/D_2/T_2$ | 2.44% | 2.04% | unknown | 26.67% | 2.05% | 66.80% |